\documentclass[aps,onecolumn,11pt]{revtex4}
\usepackage{graphicx}
\usepackage{amssymb,amsmath,amscd,amsthm}
\usepackage{times}
\newcommand{\ba}{\begin{eqnarray}}
\newcommand{\ea}{\end{eqnarray}}
\newcommand{\bsub}{\begin{subequations}}
\newcommand{\esub}{\end{subequations}}

\def\ket#1{|#1\rangle}

\usepackage[percent]{overpic}  
\usepackage{braket}

\begin{document}

\sloppy \raggedbottom

\title{Multiple quantum phase transitions in the Zr isotopes}

\author{N. Gavrielov$^1$}{}
\author{A. Leviatan$^1$}{}
\author{F. Iachello$^2$}{}

\address{$^1$Racah Institute of Physics, The Hebrew University,
  Jerusalem 91904, Israel}{}
\address{$^2$Center for Theoretical Physics, Sloane Physics Laboratory,
  Yale University, New Haven, Connecticut 06520-8120, USA}{}


\begin{abstract}
  We present a detailed analysis of spectra and other observables
  for the entire chain of Zr isotopes, from neutron number 52 to 70,
  in the framework of the interacting boson model with configuration
  mixing. The results suggest a remarkable interplay of multiple
  quantum phase transitions (QPTs). One type of QPT involves an abrupt
  crossing of normal and intruder configurations,
  superimposed on a second type of QPT involving gradual
  shape-changes within each configuration.
\end{abstract}
\maketitle
\hspace{0.6cm}

\hspace{0.6cm}
model with configuration mixing, Zr isotopes.

\section{Introduction}
Quantum phase transitions (QPTs) are qualitative changes in the structure
of a physical system induced by a change in one or more parameters that
appear in the quantum Hamiltonian describing the system.
Such ground-state phase transitions~\cite{feng,gilmore}
have been the subject of many investigations in nuclear 
physics~\cite{jolie,cejnar,iachello}, where most of the attention
has been devoted to shape phase transitions in a single configuration,
described by a single Hamiltonian of the form,
\ba
\label{eq:type-I}
\hat{H} = \left( 1-\xi \right) \hat{H}_{1}+\xi \hat{H}_{2} ~.
\ea
As the control parameter $\xi $ changes from $0$ to $1$, 
the symmetry and equilibrium shape of the system change from those of 
$\hat{H}_{1}$ to those of $\hat{H}_{2} $. For sake of clarity, 
we denote these phase transitions Type~I.
The latter have been observed
in the neutron number 90 region, {\it e.g.}, for
Nd-Sm-Gd isotopes~\cite{cejnar}.

A different type of phase transitions occurs when two (or more)
configurations coexist~\cite{Heyde11}.
In this case, the quantum Hamiltonian has a matrix form~\cite{frank}
\ba
\hat{H} &=&
\left [
\begin{array}{cc}
\hat{H}_{A}(\xi_A) & \hat{W}(\omega) \\ 
\hat{W}(\omega) & \hat{H}_{B}(\xi_B)
\end{array}
\right ] ~,\quad
\label{Hmat}
\ea
where the index $A$, $B$ denotes the two configurations 
and $\hat{W}$ denotes their coupling. We call for sake of
clarity these phase transitions Type~II,
to distinguish them from those of a single configuration.
The latter have been observed in nuclei near shell closure,
{\it e.g.}, in the light Pb-Hg isotopes~\cite{Heyde11}.

As the control parameters $(\xi_A,\xi_B,\omega)$ in Eq.~(\ref{Hmat})
are varied,
the separate Hamiltonians $\hat{H}_A$ and $\hat{H}_B$ can undergo
shape-phase transitions of Type~I, and the combined Hamiltonian
can experience a Type-II crossing of configurations $A$ and $B$.
In most cases encountered in nuclei, the separate QPTs are masked
by the strong mixing between the two configurations.
In the present contribution,
we show that the Zr isotopes are exceptional in the sense that
the crossing is abrupt, the separate configurations retain their
purity before and after the crossing, and the shape evolution
of each configuration can be cast in terms of its
own phase transition. This results in an intricate interplay
of intertwined multiple QPTs~\cite{GavLevIac19}.

\section{The IBM with configuration mixing in the Zr chain}
The $_{40}$Zr isotopes have been recently the subject of several
experimental investigations~\cite{chakraborty,Browne15,pietralla,
  Ansari17,Paul17,Witt18,Singh18}
and theoretical studies,
including mean-field based methods~\cite{delaroche,mei,nomura16},
the Monte-Carlo shell-model (MCSM)~\cite{taka} and algebraic
methods~\cite{GavLevIac19,RamHeyde19}.
We adapt here the algebraic approach of the
Interacting Boson Model (IBM)~\cite{ibm},
with bosons representing valence nucleon pairs counted from the
nearest closed shells. This provides a simple tractable framework,
where phases of quadrupole shapes: spherical, prolate-deformed and
$\gamma$-unstable deformed, correspond to U(5), SU(3) and SO(6)
dynamical symmetries, respectively.

To be specific, we use the configuration mixing model
(IBM-CM) of~\cite{duval81}, and
write the Hamiltonian not in matrix form, but rather in the equivalent form
${\textstyle\hat{H}\!=\!\hat{H}_{A}^{(N)} + \hat{H}_{B}^{(N+2)}
  + \hat{W}^{(N,N+2)}}$,
where $\hat{\cal O}^{(N)}\!=\!\hat{P}_{N}^{\dag }\hat{\cal O}\hat{P}_{N}$ 
and $\hat{\cal O}^{(N,N^{\prime })} \!=\!
\hat{P}_{N}^{\dag }\hat{\cal O}\hat{P}_{N^{\prime }}$, 
for an operator $\hat{\cal O}$, with $\hat{P}_{N}$, a projection operator 
onto the $[N] $ boson space. Here 
$\hat{H}_{A}^{(N)}$ represents the so-called normal 
($N$ boson space) configuration and $\hat{H}_{B}^{(N+2)}$
represents the so-called intruder ($N\!+\!2$ boson space) configuration.
Similar to a calculation done for the $_{42}$Mo isotopes
in~\cite{sambataro}, we consider
$_{40}^{90}$Zr$_{50}$ as a core and valence neutrons in the 50-82 major
shell. The normal $A$-configuration corresponds to having
no active protons above the $Z\!=\!40$ sub-shell gap,
and the intruder $B$-configuration corresponds to two-proton excitation
from below to above this gap, creating 2p-2h states
(see Figure~1 of~\cite{sambataro}).
The explicit form of the Hamiltonians employed in the current study is
\begin{subequations}
\label{Hmat-Zr}
\begin{align}
\hat{H}_{A} =&\,\epsilon _{d}^{(A)}\hat{n}_{d}
+\kappa ^{(A)}\hat{Q}_{\chi}\cdot \hat{Q}_{\chi } ~,
\label{HA}
\\
\hat{H}_{B} =&\,\epsilon _{d}^{(B)}\hat{n}_{d} 
+\kappa ^{(B)}\hat{Q}_{\chi}\cdot \hat{Q}_{\chi} 
+\kappa ^{\prime (B)}\hat{L}\cdot \hat{L} 
+ \Delta_p ~, 
\label{HB}
\\
\hat{W} =&\,\omega\,[\,( d^{\dag }\times d^{\dag })^{(0)}
+ \,(s^{\dag })^2\,] + {\rm H.c.} ~,
\label{W}
\end{align}
\end{subequations}
where H.c. stands for Hermitian conjugate.
The quadrupole operator is defined as 
$\hat{Q}_{\chi} = d^{\dag}s+s^{\dag }\tilde{d}
+\chi ( d^{\dag}\times \tilde{d}) ^{(2)}$ and
$\hat{n}_d$ is the $d$-boson number operator.
In Eq.~(\ref{HB}), $\Delta_p$ is the off-set between the normal
and intruder configurations,
where the index $p$ denotes the fact that this is a proton excitation.
Such Hamiltonians have been used extensively for studying coexistence
phenomena nuclei~\cite{duval81, sambataro,ramos11,ramos14,lev18}.
The resulting eigenstates $\ket{\Psi;L}$ 
with angular momentum $L$, are linear combinations of the
wave functions, $\Psi_A$ and $\Psi_B$,
in the two spaces $[N]$ and $[N+2]$,
\ba
\label{wf}
\ket{\Psi; L} = a\,\ket{\Psi_A; [N], L} + b\,\ket{\Psi_B; [N+2], L} ~,
\ea
with $a^{2}+b^{2}=1$.
\begin{figure*}[t]
\centerline{\includegraphics[width=14cm]{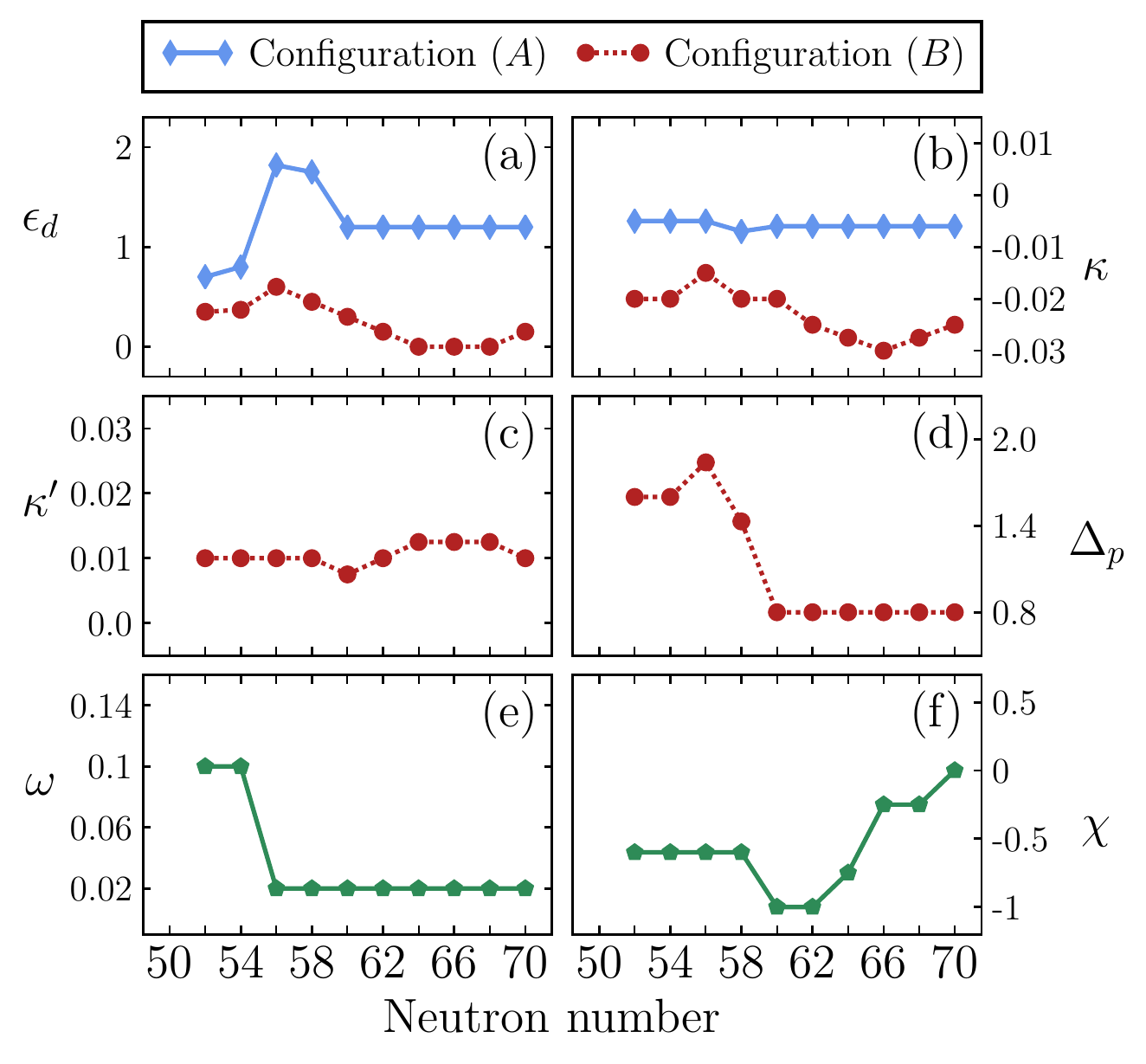}}
\caption[]{
  Parameters of the IBM-CM Hamiltonian, Eq.~(\ref{Hmat-Zr}),
  are in MeV, except for the parameter $\chi$ which is dimensionless.
Based on Table~I of~\cite{GavLevIac19}.}
\label{fig:params}
\end{figure*}

By employing the IBM-CM framework described above,
we have calculated the spectra and other observables
of the entire chain of Zr isotopes, from neutron number 52 to 70. 
The values of the Hamiltonian parameters, obtained by a global fit
to energy and $E2$ data, are shown in Figure~\ref{fig:params}, and are
consistent with those of previous calculations in this
mass region~\cite{sambataro}.
It should be noted that beyond the middle of the 
shell, at neutron number 66, bosons are replaced by boson holes~\cite{ibm}.
Apart from some fluctuations due to 
the subshell closure at neutron number 56
(the filling of the $2d_{5/2}$ orbital), the values of the parameters are 
a smooth function of neutron number and, in some cases, a constant.
A notable exception is the sharp decrease by 1~MeV of
the energy off-set parameter $\Delta_p$
beyond neutron number 56. Such a behavior was observed for the
Mo and Ge chains and, as noted in~\cite{sambataro}, it reflects the
effects of the isoscalar
residual interaction between protons and neutrons occupying the
partner orbitals $1g_{9/2}$ and $1g_{7/2}$,
which is the established mechanism for descending cross shell-gap
excitations and onset of deformation in this
region~\cite{FedPit79,HeyCas85,Heyde87}.
The $E2$ operator reads 
$\hat{T}(E2)\!=\!e^{(A)}\hat Q^{(N)}_{\chi}+e^{(B)}\hat Q^{(N+2)}_{\chi}$, 
where $\hat{Q}_{\chi}^{(N)}\!=\!\hat{P}_{N}^{\dag }\hat{Q}_{\chi}\hat{P}_{N}$,
$\hat{Q}_{\chi}^{(N+2)}\!=\!P_{N+2}^{\dag }\hat{Q}_{\chi}\hat{P}_{N+2}$
and $\hat{Q}_{\chi}$ is the same operator as in the
Hamiltonian~(\ref{Hmat-Zr}). The boson effective charges are
$e^{(A)}\!=\!0.9$ and $e^{(B)}\!=\!2.24$ $({\rm W.u.})^{1/2}$.
\begin{figure}[b]
\centering
\begin{overpic}[width=15cm]{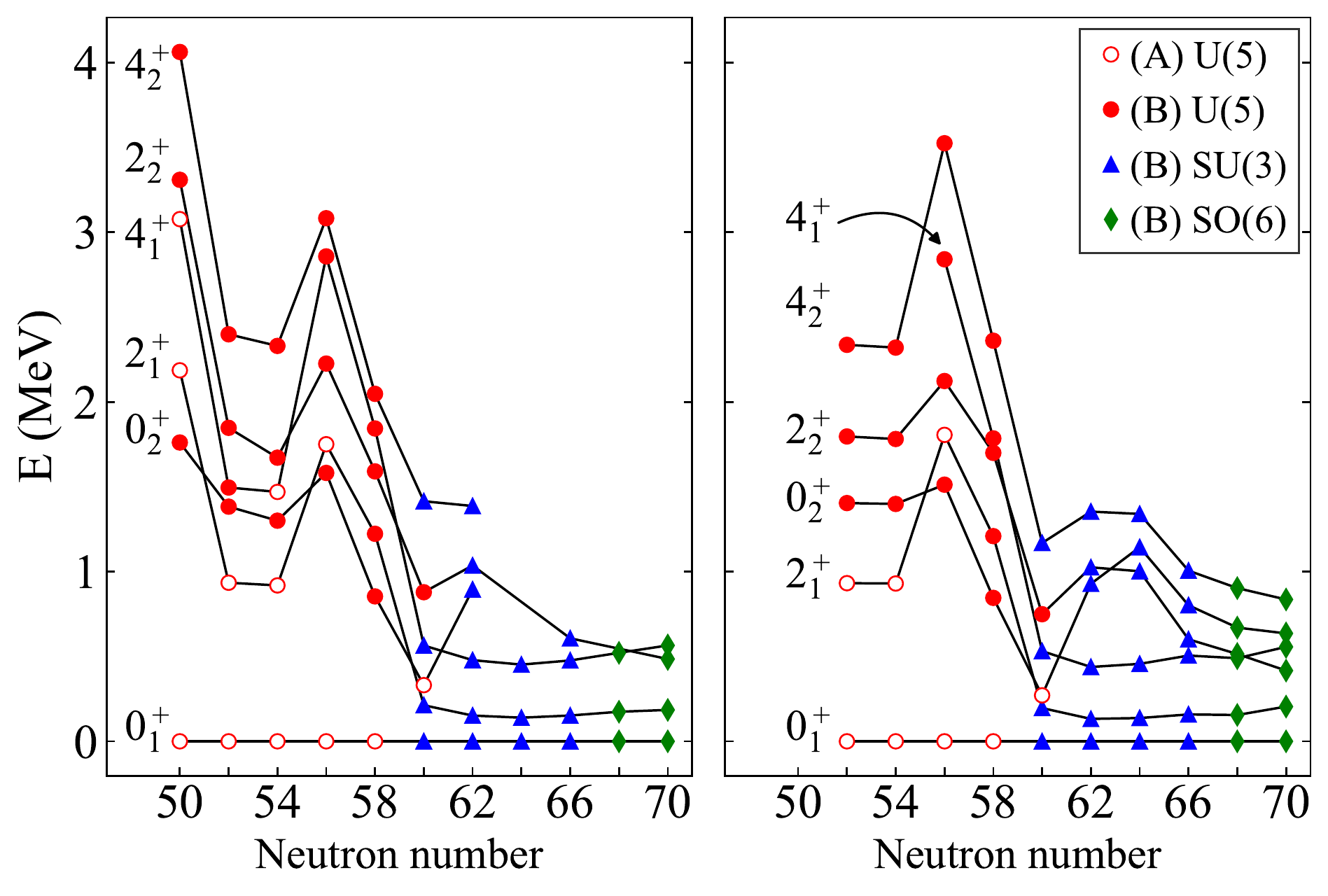}
\put (31,61) {\Large (a) {Exp}}
\put (56,61) {\Large (b) {Calc}}
\end{overpic}
\caption{Comparison between (a)~experimental~\cite{Paul17,ensdf} and
(b)~calculated energy levels
$0_{1}^{+},2_{1}^{+},4_{1}^{+},0_{2}^{+},2_{2}^{+},4_{2}^{+}$.
Empty (filled) symbols indicate a state dominated by the
normal $A$-configuration (intruder $B$-configuration),
with assignments based on the decomposition of Eq.~(\ref{wf}).
The shape of the symbol [$\circ,\,\triangle,\,\Diamond$], 
indicates the closest dynamical symmetry
[U(5), SU(3), SO(6)] to the level considered.
Note that the calculated values
start at neutron number 52, while the experimental values include the
closed shell at 50. Adapted from~\cite{GavLevIac19}.
\label{fig:levels}}
\end{figure}
\begin{figure*}[t]
\centering
\includegraphics[width=7.5cm]{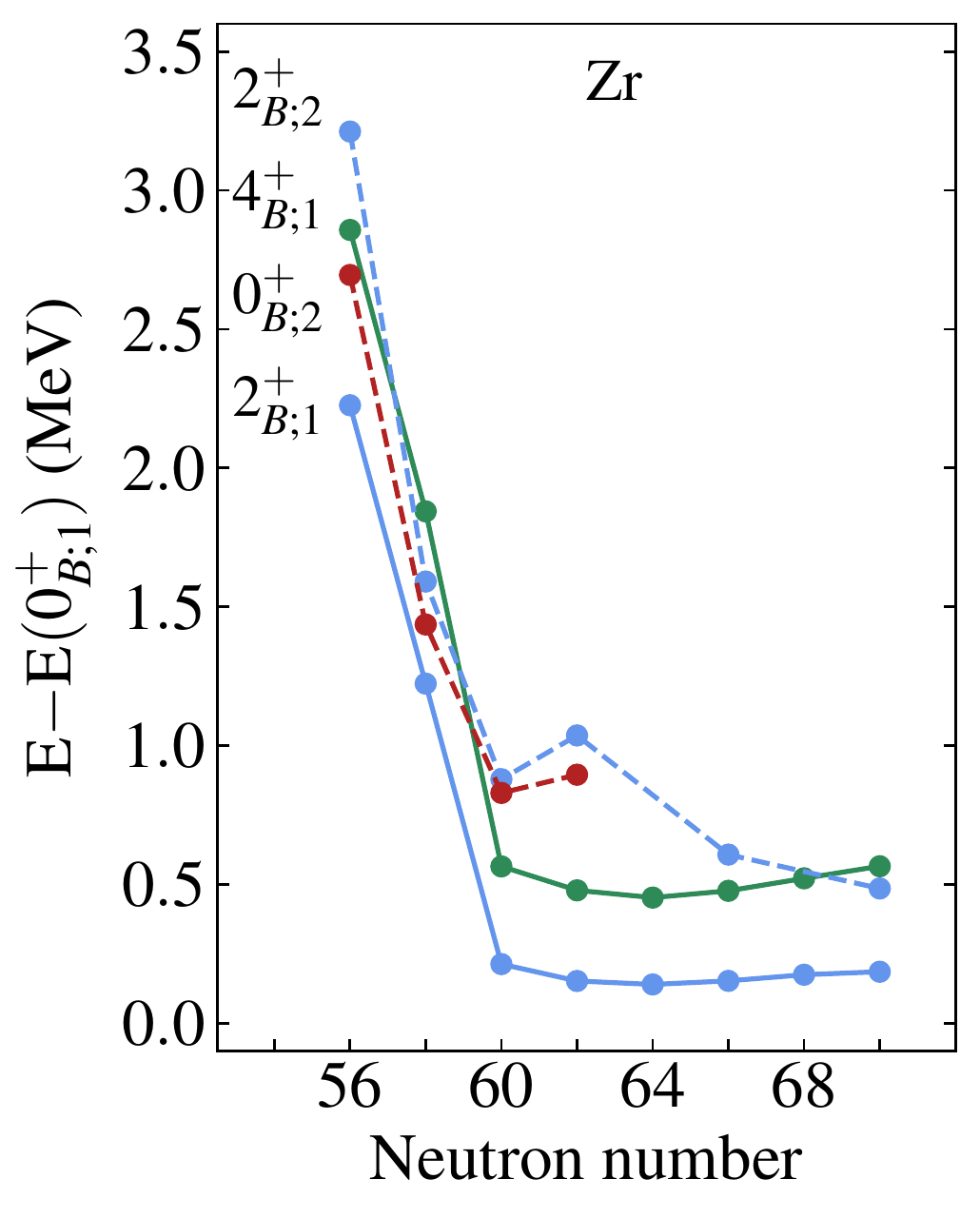}
\includegraphics[width=7.5cm]{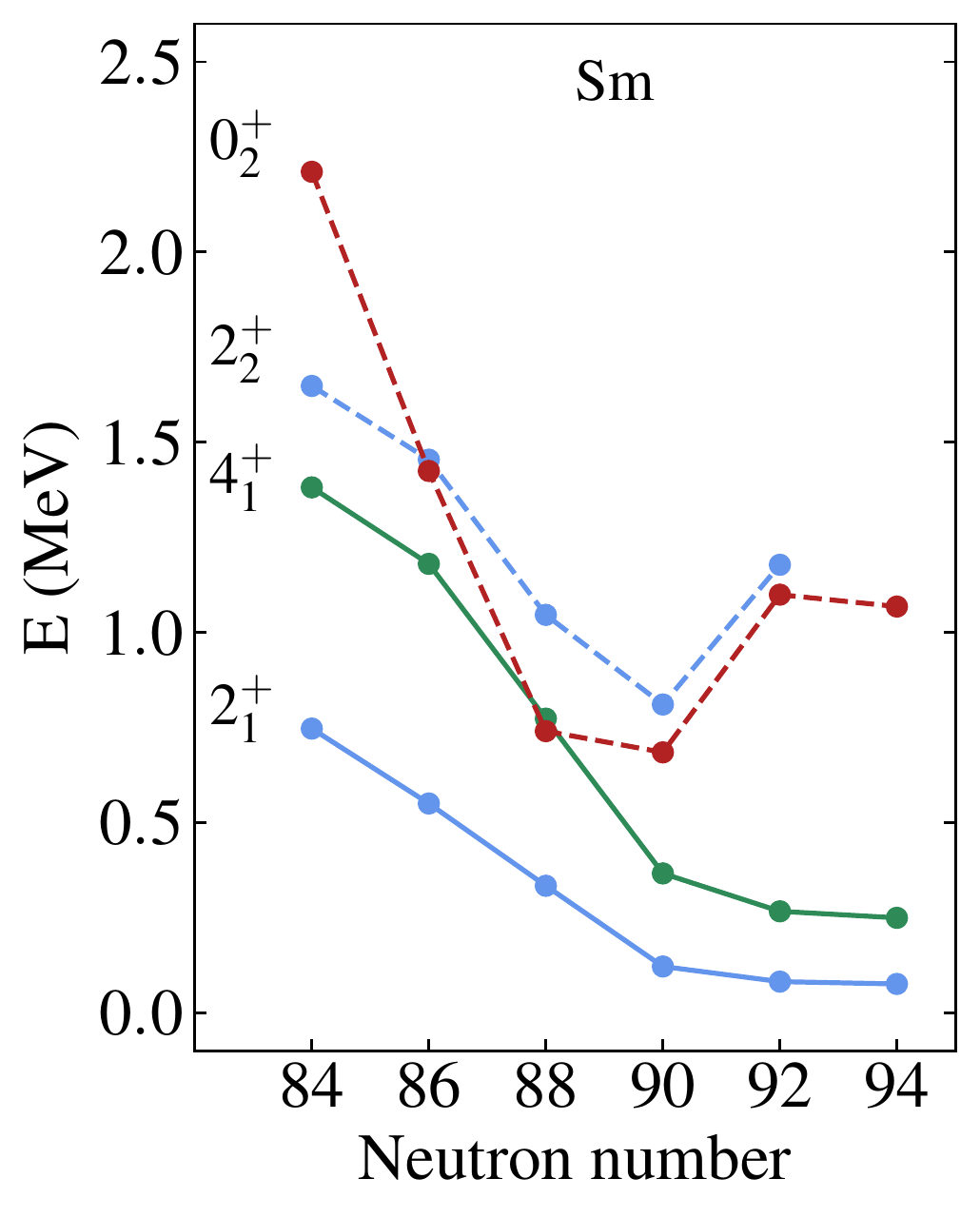}
\caption[]{Left panel: Experimental excitation energies for states of
  the intruder $B$-configuration (denoted by $L^{+}_{B;i}$ ) 
  in the $_{40}$Zr isotopes, with respect to the lowest state ($0^{+}_{B;1}$)
  of that configuration.
  Right panel:
  Typical features of a U(5)-SU(3) Type~I QPT manifested in the
  empirical spectra of the $_{62}$Sm isotopes. Data taken from~\cite{ensdf}.}
\label{Zr-Sm}
\end{figure*}

\section{Evolution of spectra along the Zr chain}
An important clue for understanding the change in structure of the Zr
isotopes, is obtained by examining the evolution of their spectra along
the chain.
In Figure~\ref{fig:levels}, we show a comparison between experimental and 
calculated levels, together with assignments to configurations based on
Eq.~(\ref{wf}), and to the closest dynamical symmetry for each level.
One can see here a rather complex structure.
In the region between neutron number 50 and 56, there appear to be two
configurations, one spherical (seniority-like), ($A$),
and one weakly deformed, ($B$), as evidenced
by the ratio $R_{4/2}$ in each configuration
which is at 52-56, $R^{(A)}_{4/2}\cong 1.6 $
and $R^{(B)}_{4/2} \cong 2.3$.
From neutron number 58, there is a pronounced drop in energy for the
states of configuration~($B$), and at 60, the two configurations
exchange their roles, indicating a Type~II QPT.
At this stage,
the intruder configuration~($B$) appears to be at the critical point
of a U(5)-SU(3) Type~I QPT, as evidenced
by the low value of the excitation energy of the
first-excited $0^+$ state of the $B$ configuration,
(denoted by $0^{+}_{B;2}$ in the left panel of Figure~\ref{Zr-Sm}).
The same situation occurs in the $_{62}$Sm and $_{64}$Gd
isotopes at neutron number 90~\cite{ibm,scholten78},
as is clearly shown in Figure~\ref{Zr-Sm}.
Beyond neutron number 60, the intruder configuration~($B$)
becomes progressively strongly deformed, as evidenced by the small
value of the excitation energy of the state $2_{1}^{+}$,
$E_{2^+_1}\!=\!151.78$ keV and by the ratio $R^{(B)}_{4/2}\!=\!3.15$
in $^{102}$Zr, and $E_{2_{1}^{+}}\!=\!139.3$~keV, $R^{(B)}_{4/2}\!=\!3.24$
in $^{104}$Zr. At still larger neutron number 66,
the ground state band becomes $\gamma $-unstable,
as evidenced by the close energy of the states $2_{2}^{+}$ and $4_{1}^{+}$, 
$E_{2_{2}^{+}}\!=\!607.0$~keV, $E_{4_{1}^{+}}\!=\!476.5$ keV, in $^{106}$Zr, 
and especially by the recent results 
$ E_{4^+_1}\!=\!565$~keV and $ E_{2^+_2}\!=\!485$ keV 
in $^{110} $Zr~\cite{Paul17}, a signature of the SO(6) symmetry. 
In this region, the ground state 
configuration undergoes a~crossover from SU(3) to SO(6).

\section{Type I and Type II QPTs in the Zr chain}
The above spectral analysis signal the presence of multiple QPTs
in the Zr isotopes.
In order to understand the nature of these phase
transitions, it is instructive to examine in more detail
the way in which the structure of the states evolves, and the
behaviour of the order parameters.

\begin{figure*}[b]
  \centerline{\includegraphics[width=14cm]{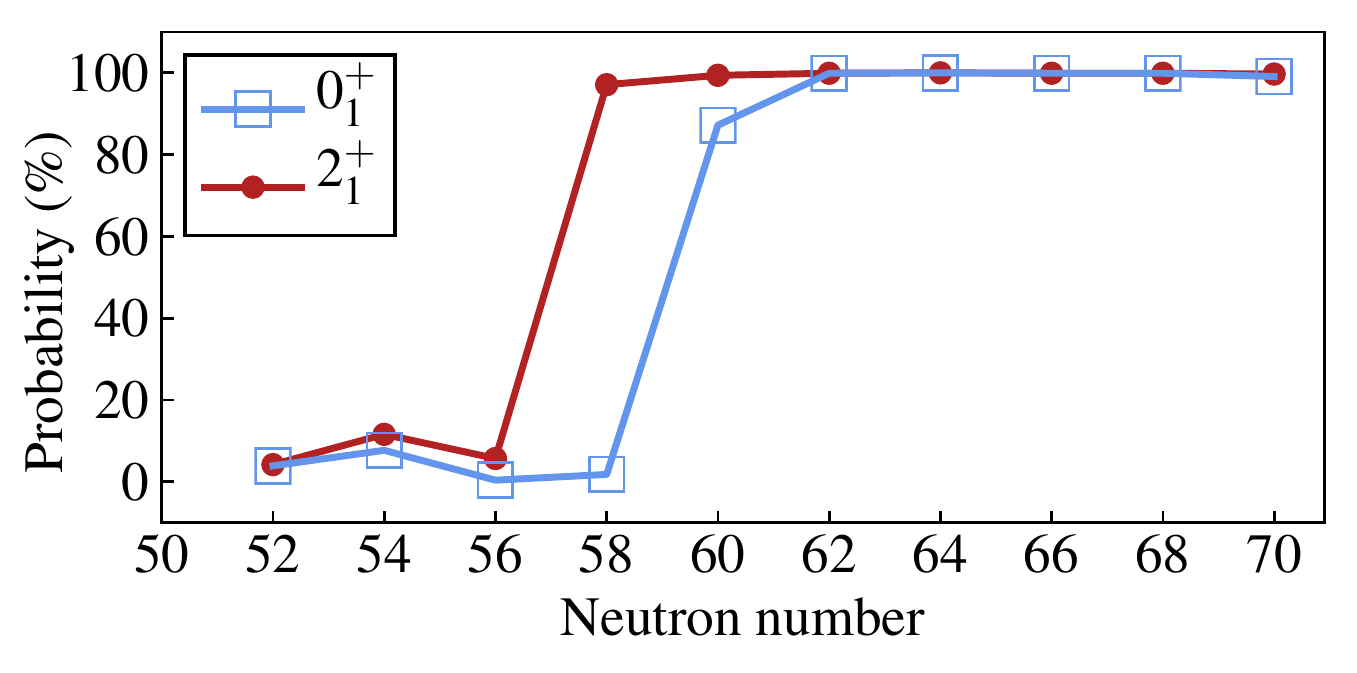}}
  \caption[]{Percentage of the wave functions within the intruder
  $B$-configuration [the $b^2$ probability in Eq.~(\ref{wf})],
  for the ground ($0^{+}_1$) and excited ($2^{+}_1$) states in
  each Zr isotope.
\label{fig:decomp}}
\end{figure*}

Information on configuration-changes
can be inferred from the probabilities $a^2$ or $b^2$, Eq.~(\ref{wf}),
of the states considered.
Figure~\ref{fig:decomp} shows the  percentage of the wave function
within the intruder configuration for the ground state
($0^{+}_1$) and first-excited state ($2^{+}_1$) as a function of neutron
number along the Zr chain.
The rapid change in structure of the $0^{+}_1$ state from the
normal $A$-configuration $(b^2 \!=\! 1.8\%)$ in $^{98}$Zr
to the intruder $B$-configuration ($b^2 \!=\! 87.2\%)$
in $^{100}$Zr, is clearly evident.
The jump in configurations appears sooner in the
$2^+_1$ state, which changes to configuration~$B$ ($b^2 \!=\! 97.1\%$)
already in $^{98}$Zr, as pointed out in~\cite{Witt18}.
Outside a narrow region near neutron number 60 where the crossing occurs,
the two configurations are weakly mixed and the states
retain a high level of purity ({\it i.e.}, $b^2\approx 0\%$ or $100\%$).

\begin{figure*}[t]
  \centerline{\includegraphics[width=12cm]{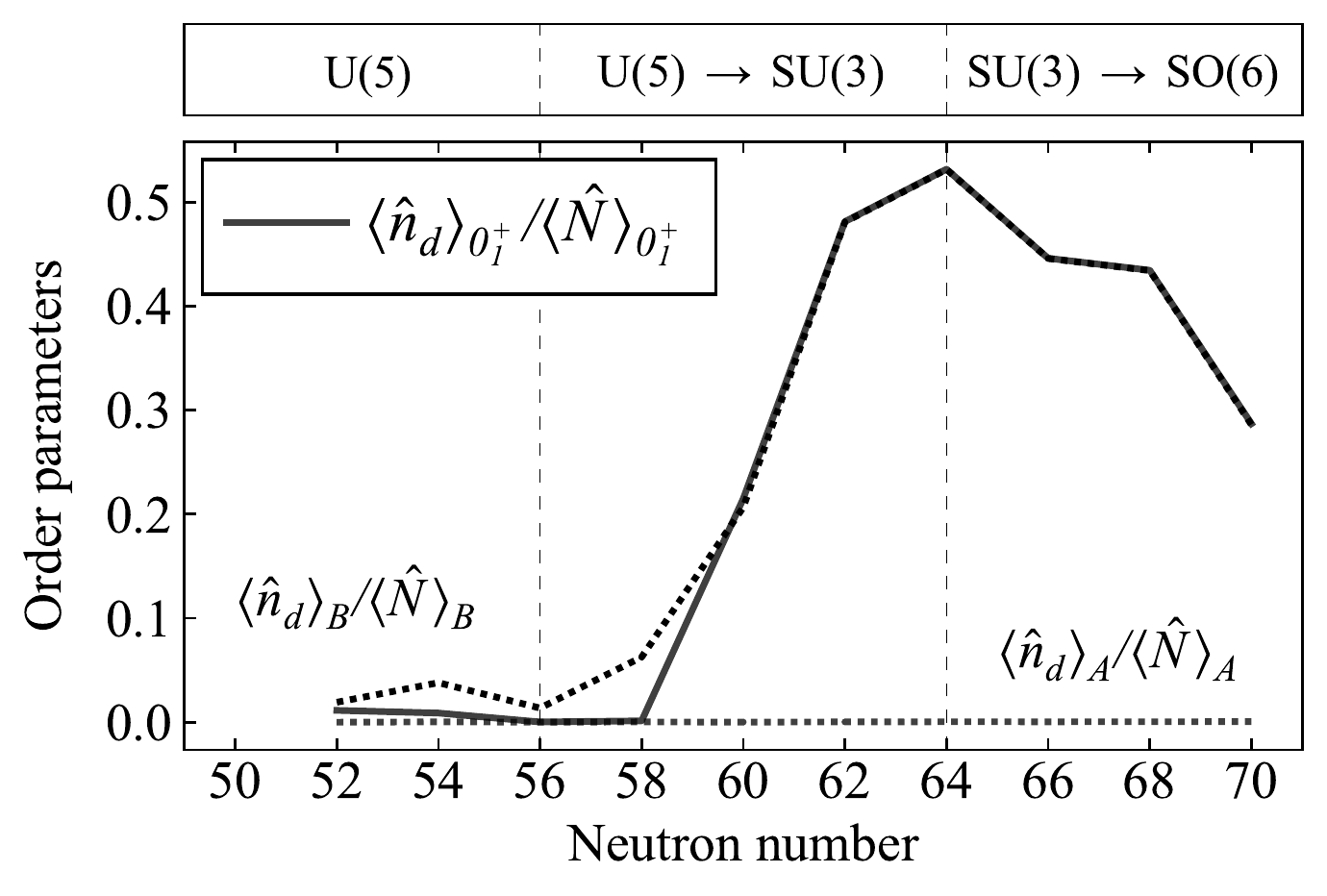}}
  \caption[]{Evolution of order parameters along the
Zr chain. The latter are the calculated
expectation values of $\hat{n}_d$ in the
total ground state wave function $\ket{\Psi; L=0^{+}_1}$,
Eq.~(\ref{wf}) (solid line) and in its ($A$) and ($B$) components
(dotted lines), normalized by the respective boson numbers
$\braket{\hat{N}}_{0^{+}_1}\!=\!a^2N\!+\!b^2(N\!+\!2)$,
$\braket{\hat{N}}_A\!=\!N$, $\braket{\hat{N}}_B\!=\!N\!+\!2$.
Adapted from~\cite{GavLevIac19}.}
\label{order-param}
\end{figure*}

Information on shape-changes can be obtained by examining the relevant
order parameters of the phase transitions.
In the present study, the latter involve
the expectation value of $\hat{n}_d$ in the ground state wave function,
$\ket{\Psi; L\!=\!0^{+}_1}$ and in its
$\Psi_A$ and $\Psi_B$ components~(\ref{wf}),
denoted by $\braket{\hat{n}_d}_{0^{+}_1}$, $\braket{\hat{n}_d}_A$,
and $\braket{\hat{n}_d}_B$.
Here $\braket{\hat{n}_d}_A$ and $\braket{\hat{n}_d}_B$ portray 
the shape of configuration~($A$) and ($B$), respectively, 
and $\braket{\hat{n}_d}_{0^{+}_1}= a^2\braket{\hat{n}_d}_A
+ b^2\braket{\hat{n}_d}_B$
portrays also the dependence on the normal-intruder mixing.
Figure~\ref{order-param} shows the evolution along the Zr chain 
of these order parameters 
($\braket{\hat{n}_d}_{A},\,\braket{\hat{n}_d}_{B}$ in dotted 
and $\braket{\hat{n}_d}_{0^{+}_1}$ in solid lines),
normalized by the respective boson numbers.
Configuration~($A$) is seen to be spherical for all
neutron numbers considered. 
In contrast, configuration~($B$) is weakly-deformed
for neutron number 52-58. One can see here clearly the imprints of
the jump, noted in Figure~\ref{fig:decomp} between neutron number 58 and 60,
from configuration~($A$) to configuration~($B$), indicating
a first-order Type~II QPT, a further increase at neutron numbers 60-64
indicating a U(5)-SU(3) Type~I QPT within configuration ($B$),
and, finally, there is a decrease at neutron number 66, due in part to
the crossover from SU(3) to SO(6) and in part to the shift from boson 
particles to boson holes after the middle of the major shell 50-82. 
$\braket{\hat{n}_d}_{0^{+}_1}$ is close to $\braket{\hat{n}_d}_A$
for neutron number 52-58 and coincides with $\braket{\hat{n}_d}_B$ 
at 60 and above, consistent with a high degree of purity with respect 
to configuration-mixing.

The combined analyses of energy spectra (Figure~\ref{fig:levels}),
wave-functions (Figure~\ref{fig:decomp})
and order parameters (Figure~\ref{order-param}),
suggests a remarkable interplay of
configurations-interchange and shape-evolution in the Zr isotopes,
manifesting simultaneously multiple QPTs
with different character. Specifically,
a Type~II QPT involves an abrupt crossing of the normal and
intruder configurations. This local effect is superimposed on 
a Type~I QPT which involves a gradual shape-change within each configuration.
In particular, the normal $A$-configuration remains spherical along
the Zr chain, while the intruder $B$-configuration undergoes
a first-order U(5) $\to$ SU(3) (spherical to prolate-deformed) transition
and an SU(3) $\to$ SO(6) (prolate to $\gamma$-unstable deformed ) crossover.

\section{Empirical signatures}
Further evidence for the scenario of multiple QPTs in the Zr isotopes,
is obtained by an analysis of other observables, such as $E2$ transition
rates, isotope shifts and two-neutron separation energies.

\begin{figure*}[t]
\centerline{\includegraphics[width=11.5cm]{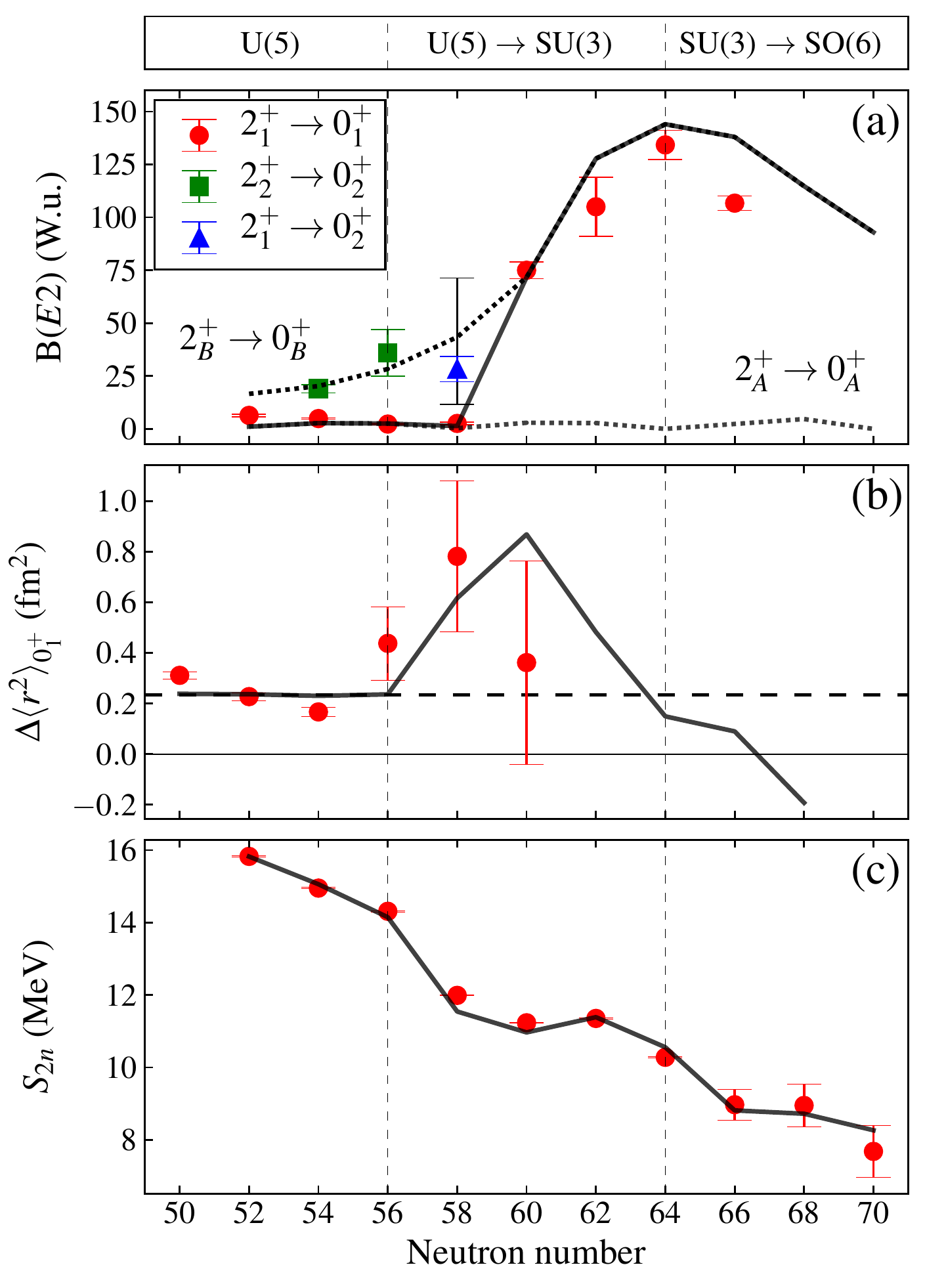}}
\caption[]{Empirical signatures for QPTs in the Zr chain.
Symbols (solid lines) denote experimental data (calculated results).
(a)~$B(E2)$ values in Weisskopf units (W.u.).
Data taken from~\cite{ensdf,chakraborty,Browne15,pietralla,Ansari17,
  Witt18,Singh18}.
Dotted lines denote calculated $E2$ transitions within a configuration.
(b)~Isotope shift, $\Delta\braket{\hat{r}^{2}}_{0^{+}_1}$
in fm$^{2}$. Data taken from~\cite{angeli2}. The horizontal dashed
line at $0.235$ fm$^{2}$ represents the smooth behavior in
$\Delta \braket{\hat{r}^{2}}_{0^{+}_1}$ due to the $A^{1/3}$ increase of
the nuclear radius. (c)~Two-neutron separation energies,
$\rm{S}_{2n}$, in MeV. Data taken from AME2016~\cite{wang-masses}.
Adapted from~\cite{GavLevIac19}.}
\label{be2-iso-s2n}
\end{figure*}

As shown in Figure~\ref{be2-iso-s2n}(a), the calculated $B(E2)$'s 
agree with the empirical values and follow the same trends as the 
respective order parameters. The dotted lines denote calculated
$E2$ transitions between states within the same configuration.
The calculated $2^{+}_A\to 0^{+}_A$ transition rates
coincide with the empirical $2^{+}_1\to~0^{+}_1$
rates for neutron number 52-56. The calculated
$2^{+}_B\to 0^{+}_B$ transition rates
coincide with the empirical $2^{+}_2\to 0^{+}_2$ rates for
neutron number 52-56, with the empirical $2^{+}_1\to 0^{+}_2$ rates
at neutron number 58, and with the empirical $2^{+}_1\to 0^{+}_1$ rates
at neutron number 60-64. The large jump in $B(E2;2^+_1\rightarrow0^+_1)$
between neutron number 58 and 60 reflects the passing through a critical
point, common to a Type~II QPT involving a crossing of two configurations
and a spherical to deformed Type~I QPT within the
$B$ configuration. The further increase in $B(E2;2^+_1\rightarrow0^+_1)$
for neutron numbers 60-64 is as expected for such 
a U(5)-SU(3) QPT (see Figure~2.20 in~\cite{ibm}) and,
as seen in Figure~\ref{order-param}, it reflects an increase of the
deformation in a spherical to deformed shape-phase transition.
The subsequent decrease from the peak at neutron number 64 towards 70,
is in accord with an SU(3) to SO(6) crossover (see Figure~2.22
in~\cite{ibm}).

Further evidence for the indicated structural changes occurring in
the Zr chain, can be obtained from an analysis of the isotope shift 
$\Delta\braket{\hat r^2}_{0^+_1}
=\braket{\hat{r}^{2}}_{0^{+}_1;A+2}-\braket{\hat{r}^2}_{0^{+}_1;A}$, where 
$\braket{\hat r^2}_{0^+_1} $ is the expectation value of $ \hat r^2 $
in the ground state $ 0^+_1 $. In the IBM-CM the latter is given by 
$\braket{\hat r^2} = r^2_c + \alpha N_v + \eta [\braket{\hat n_d^{(N)}} 
  + \braket{\hat n_d^{(N+2)}}]$,
where $r^2_c$ is the square radius of the closed shell, 
$N_v$ is half the number of valence particles, 
and $\eta$ is a coefficient that takes into account the effect 
of deformation\cite{ibm,Zerguine2008,Zerguine2012}.
The isotope shift depends on two parameters,
$\alpha\!=\!0.235,\,\eta \!=\! 0.264$ fm$^2$, whose values are fixed
by the procedure of Ref.~\cite{Zerguine2008,Zerguine2012}.
$\Delta\braket{\hat r^2}_{0^+_1}$ should increase at the transition
point and decrease and, as seen in Figure~\ref{be2-iso-s2n}(b),
it does so, although the error bars are large and no data are available
beyond neutron number 60. (In the large $N$ limit, this quantity, 
proportional to the derivative of the order parameter 
$\braket{\hat{n}_d}_{0^{+}_1}$, diverges at the critical point).

Similarly, the two-neutron separation energies $S_{2n} $ 
can be written as~\cite{ibm},
$S_{2n} = -\tilde{A} -\tilde{B} N_v \pm S^{\rm def}_{2n} - \Delta_n$,
where $S^{\rm def}_{2n}$ is the contribution of the deformation,
obtained by the expectation value of the Hamiltonian in the
ground state~$ 0^+_1$.
The $ + $ sign applies to particles and the $ - $ sign to holes,
and $\Delta_n $ takes into account the neutron subshell closure at 56, 
$\Delta_n = 0 $ for 50-56 and $ \Delta_n = 2 $ MeV for 58-70.
The value of $ \Delta_n $ is taken from Table XII of~\cite{Barea2009} 
and $ \tilde{A}\!=\!-16.5,\,\tilde{B}\!=\!0.758$ MeV are determined
by a fit to binding energies of $^{92,94,96}$Zr.
The calculated $ S_{2n}$, shown in Figure~\ref{be2-iso-s2n}(c),
displays a complex behavior. Between neutron number 52 and 56 
it is a straight line, as the ground state is spherical (seniority-like)
configuration~($A$). After 56, it first goes down due to the subshell
closure at~56, then it flattens as expected from a first-order QPT 
(see, for example the same situation in the $_{62}$Sm
isotopes~\cite{scholten78}). After 62, 
it goes down again due to the increasing of deformation and finally it
flattens as expected from a crossover from SU(3) to SO(6).

\section{Quantum and classical analyses}
One of the main advantages of the algebraic method is
that one can do both a quantum and a classical analysis.
The calculations describe the experimental data in the
entire range $^{92-110}$Zr very well.
A full account is given in~\cite{gavrielov-thes}.
Here we show
only three examples, $^{98}$Zr, $^{100}$Zr and $^{102}$Zr.

\begin{figure*}[b]
\centering
\begin{overpic}[width=0.33\linewidth,height=0.24\textheight,
keepaspectratio]{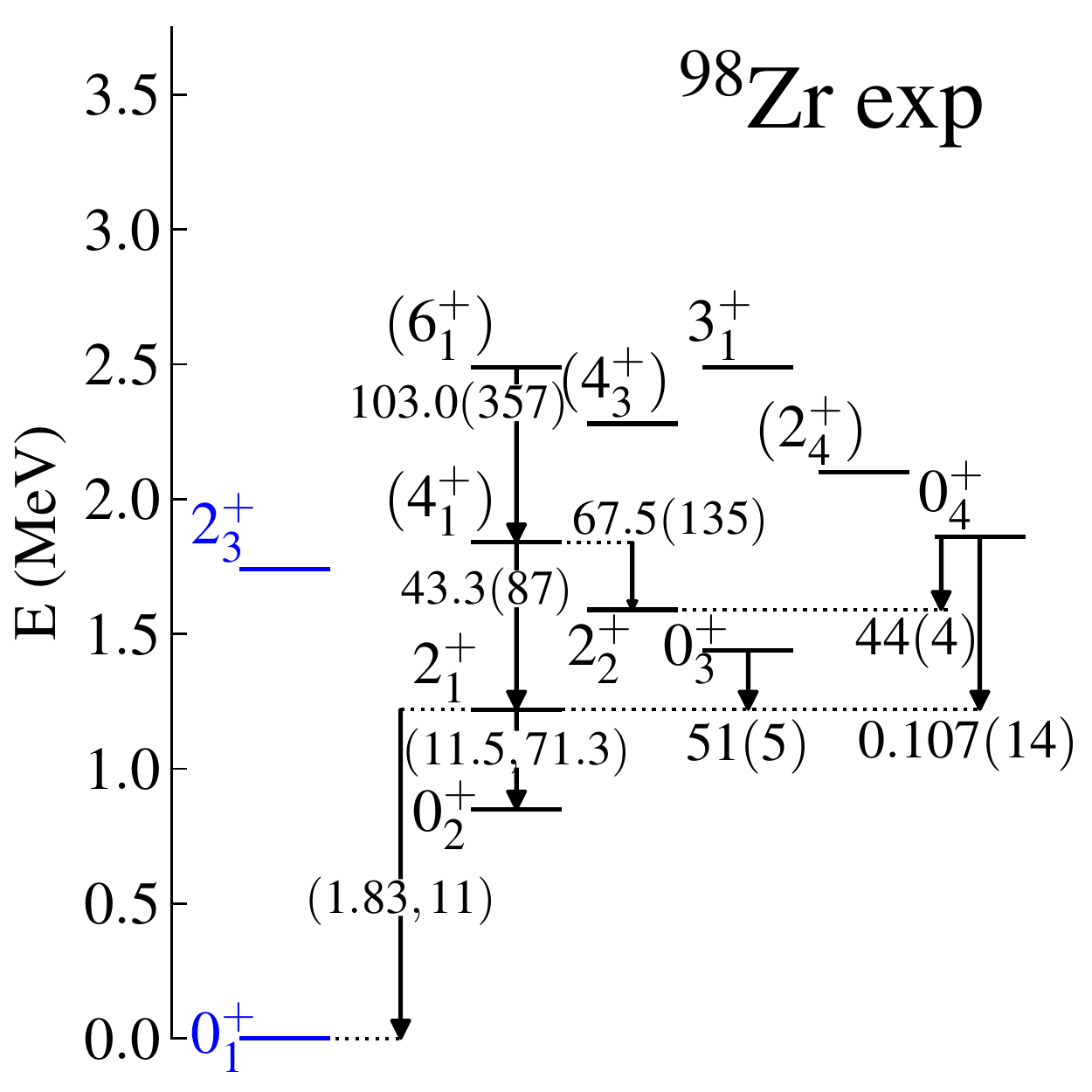}
\put (80,75) {\large(a)}
\end{overpic}
\begin{overpic}[width=0.33\linewidth,height=0.24\textheight,
keepaspectratio]{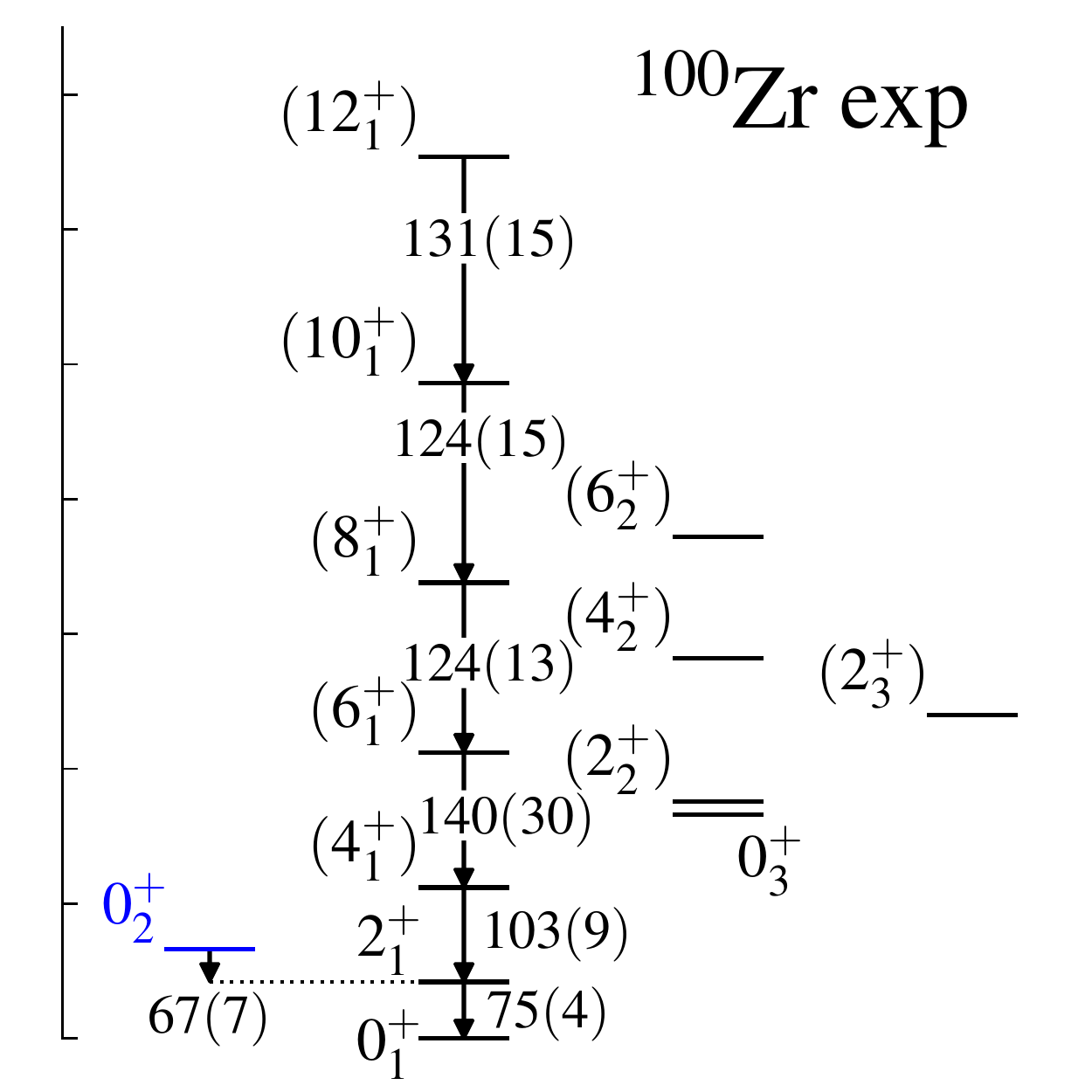}
\put (80,75) {\large(c)}
\end{overpic}
\begin{overpic}[width=0.33\linewidth,height=0.20\textheight,
keepaspectratio]{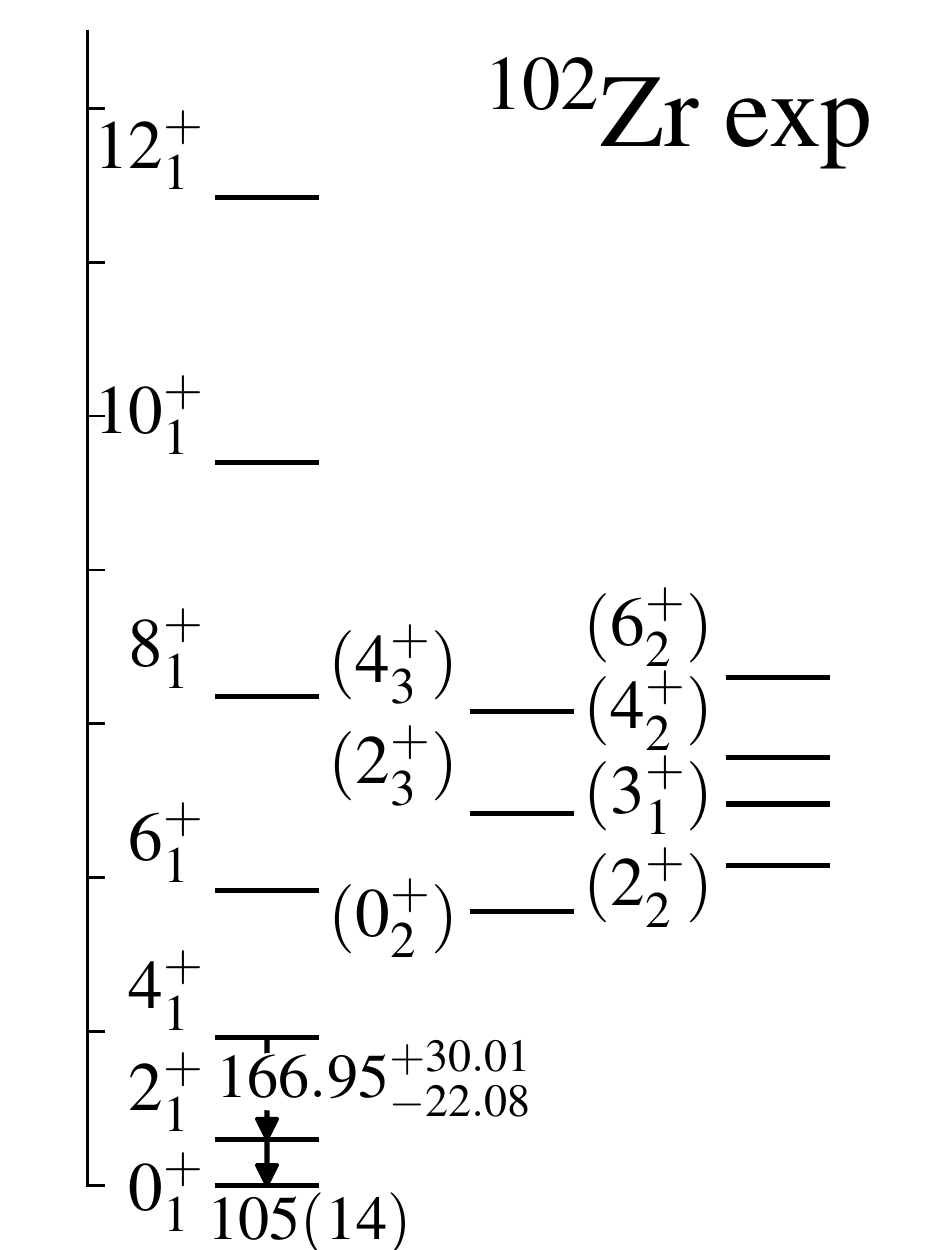}
\put (60,75) {\large(e)}
\end{overpic}\\
\begin{overpic}[width=0.33\linewidth,height=0.24\textheight,
keepaspectratio]{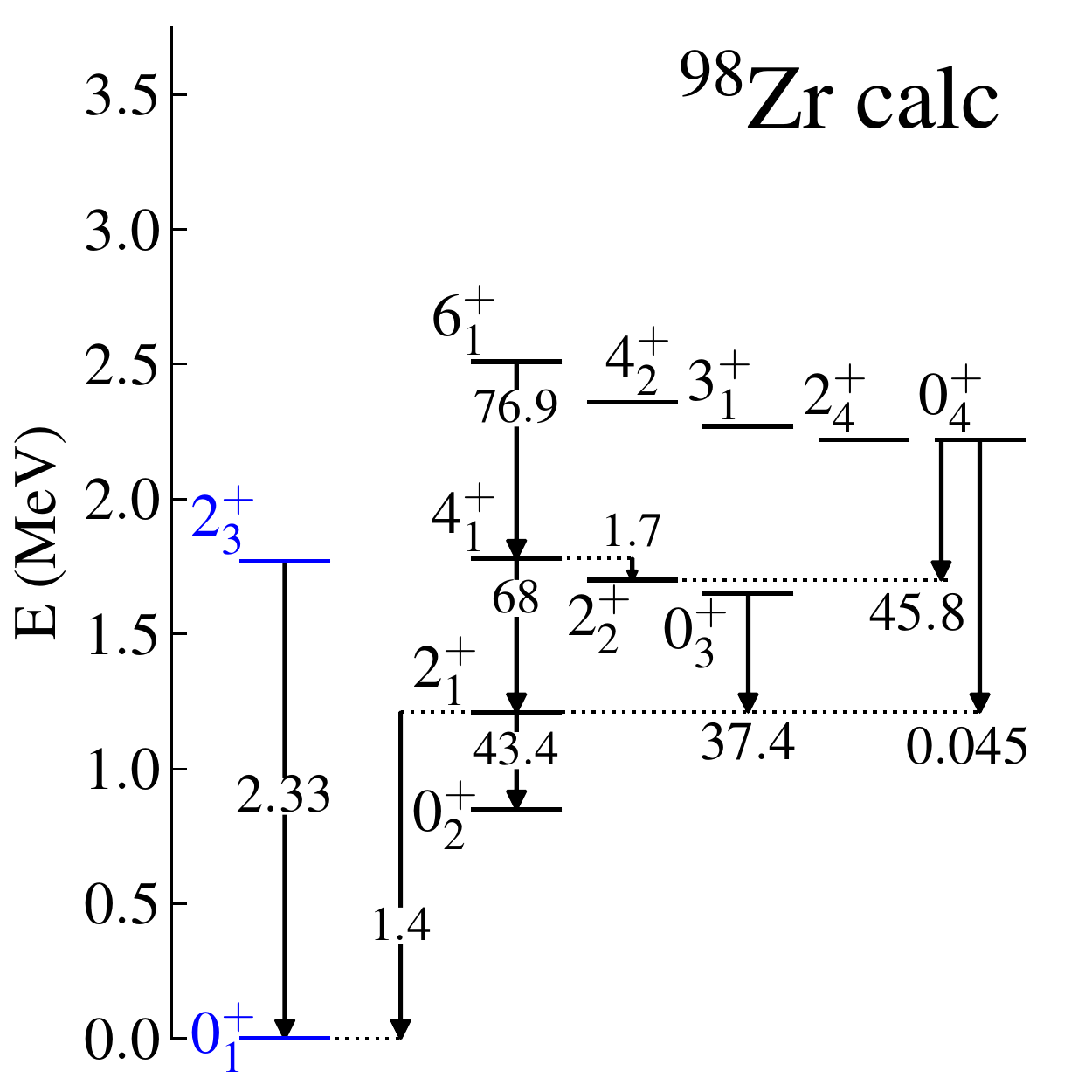}
\put (80,75) {\large(b)}
\end{overpic}
\begin{overpic}[width=0.33\linewidth,height=0.24\textheight,
keepaspectratio]{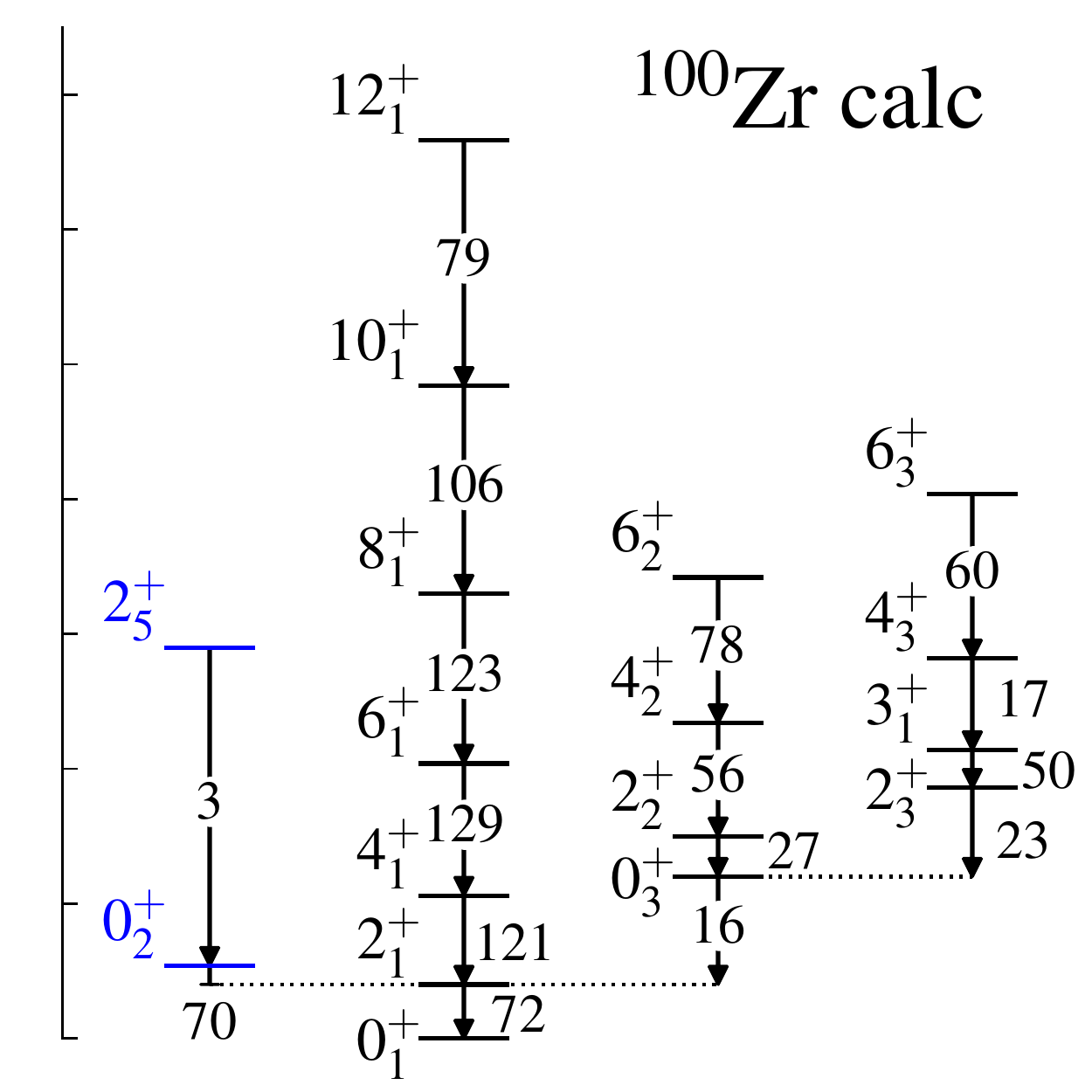}
\put (80,75) {\large(d)}
\end{overpic}
\begin{overpic}[width=0.33\linewidth,height=0.20\textheight,
keepaspectratio]{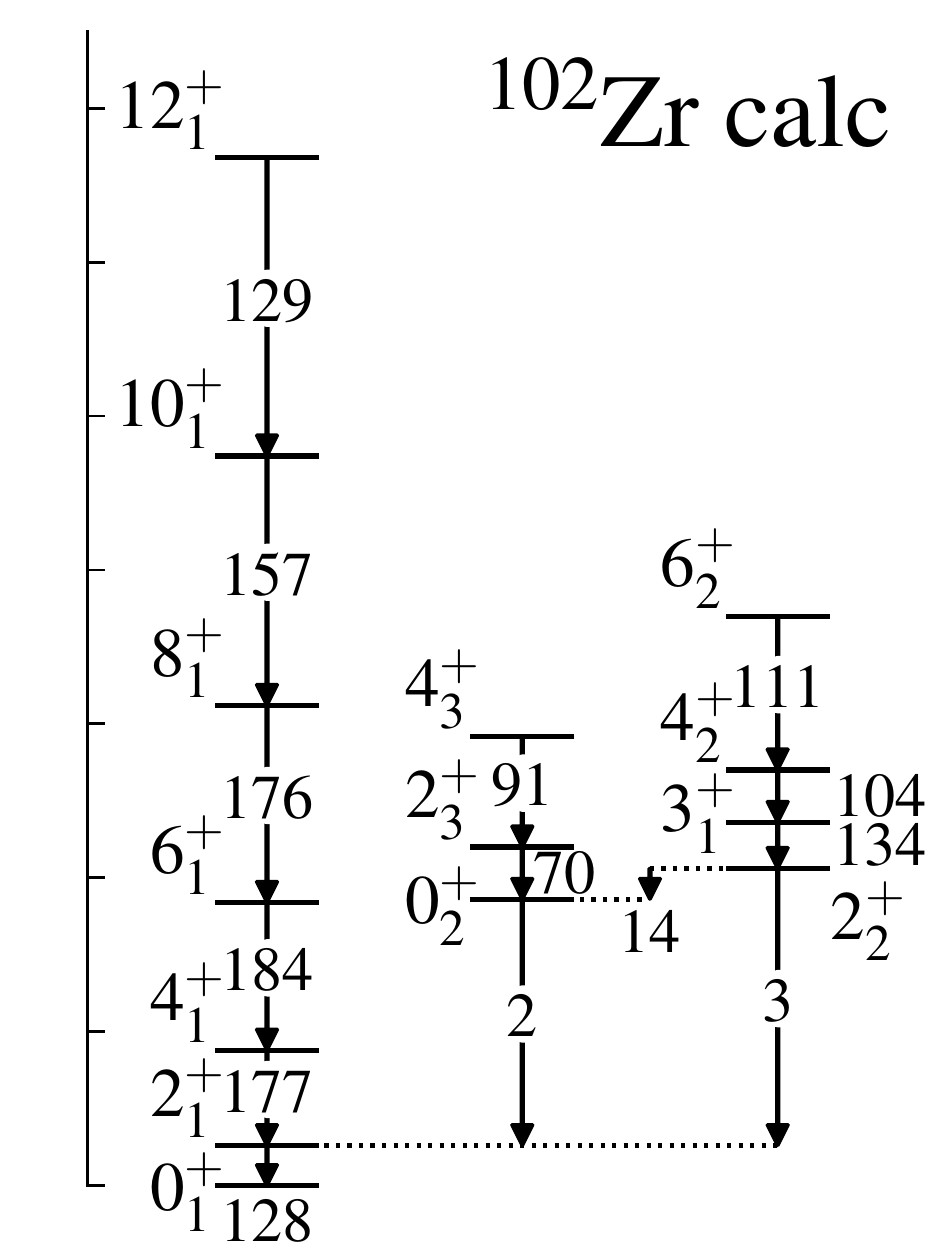}
\put (60,75) {\large(f)}
\end{overpic}
\caption{\label{fig:spectrum} \small
  Experimental~\cite{Ansari17,Witt18,Singh18,ensdf} (top row) and
  calculated (bottom row) energy levels in MeV and $E2$ rates in W.u.
  for $^{98}$Zr [panels (a)-(b)], $^{100}$Zr
  [panels (c)-(d)] and $^{102}$Zr [panels (e)-(f)].
  The levels ($0^{+}_1,\,2^{+}_3$)
  in $^{98}$Zr and ($0^{+}_2,\,2^{+}_5$) in$^{100}$Zr are dominated by
  the normal ($A$) configuration. All other levels shown are dominated by
  the intruder ($B$) configuration. Assignments are based on the
  decomposition of Eq.~(\ref{wf}). Panels (c)-(d) are adapted
  from~\cite{GavLevIac19}.}
\end{figure*}

$^{98}$Zr, in Figures~\ref{fig:spectrum}(a) and \ref{fig:spectrum}(b),
has a spherical [U(5)-like] ground state configuration ($A$)
and a weakly-deformed [U(5)-perturbed] excited configuration ($B$).
$^{100}$Zr, in Figures~\ref{fig:spectrum}(c) and \ref{fig:spectrum}(d),
is near the critical point of both
Type~I and Type~II QPTs.
The ground state band, has now become configuration~($B$), 
and appears to have features of the so-called X(5) 
symmetry~\cite{X5}, while 
the spherical configuration~($A$) has now become the excited band
$0^+_2 $. $^{102}$Zr, in Figures~\ref{fig:spectrum}(e)
and ~\ref{fig:spectrum}(f), exhibits well developed deformed
[SU(3)-like] rotational bands assigned to configuration~($B$).
States of configuration~($A$) have shifted to higher energies.

\begin{figure*}[t]
\centering
\begin{overpic}[width=0.328\linewidth]{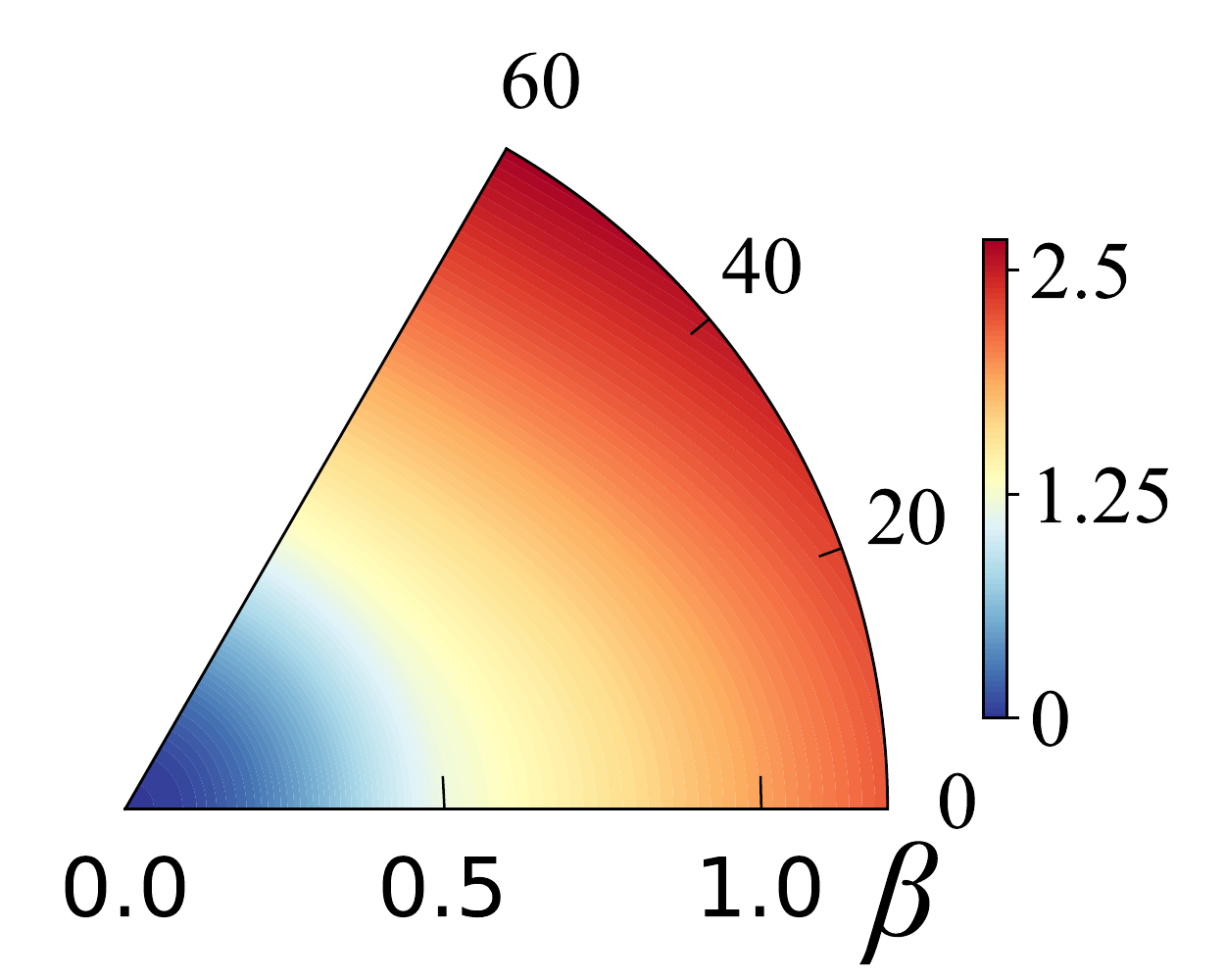}
\put (0,60) {\large $^{98}$Zr}
\end{overpic}
\begin{overpic}[width=0.328\linewidth]{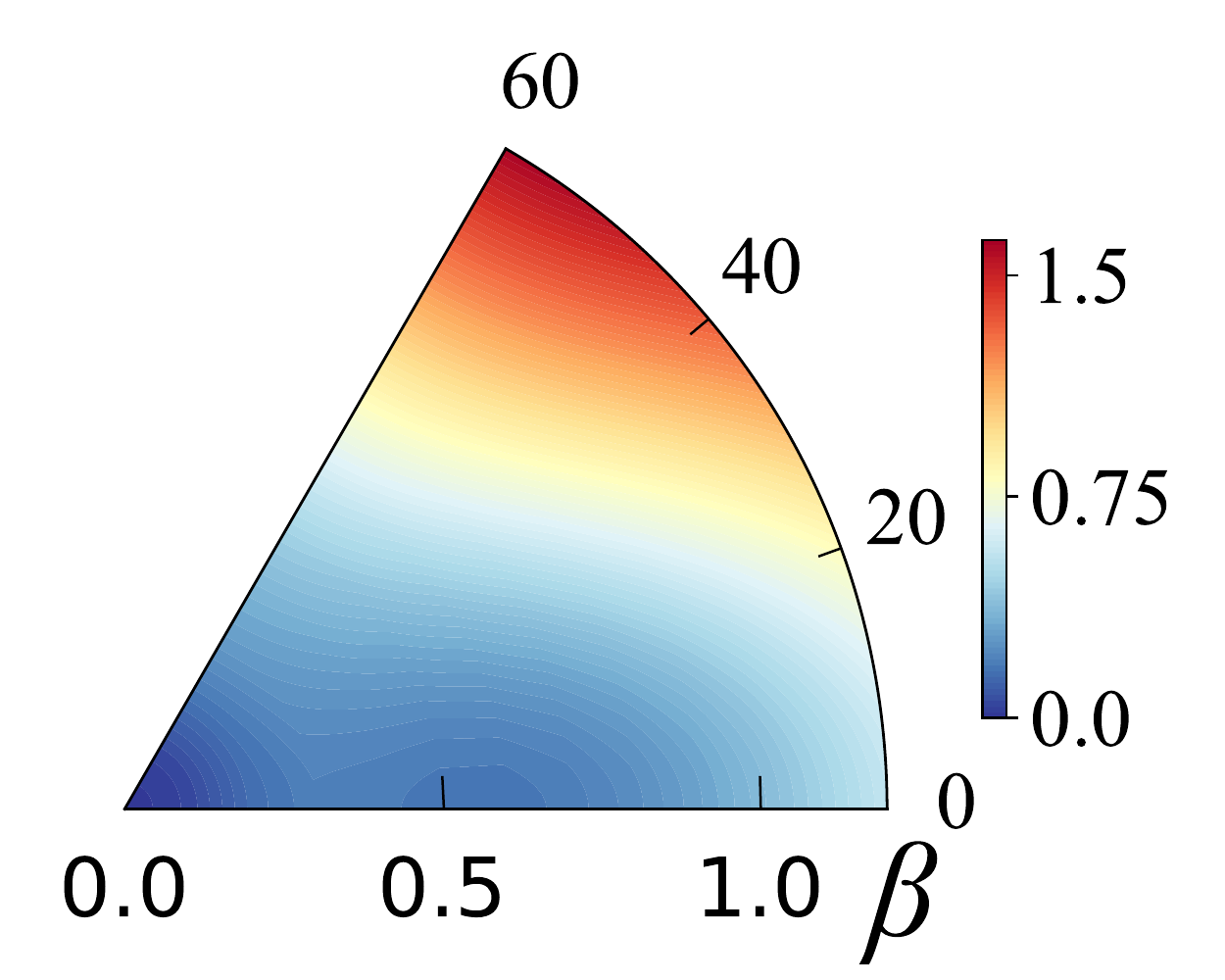}
\put (0,60) {\large $^{100}$Zr}
\end{overpic}
\begin{overpic}[width=0.328\linewidth]{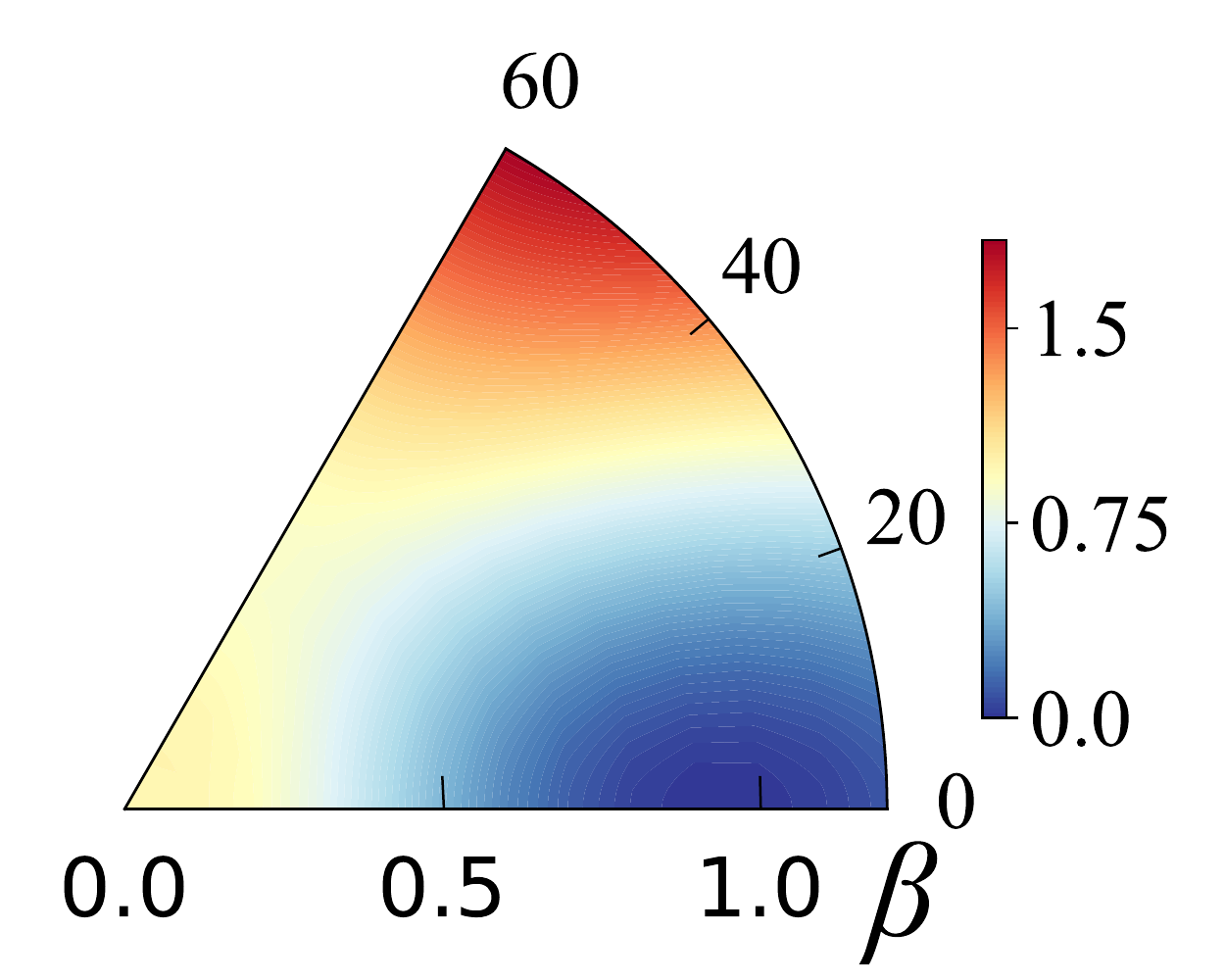}
\put (0,60) {\large $^{102}$Zr}
\end{overpic}
\caption{\label{Eminus}
\small
Contour plots in the $(\beta ,\gamma )$ plane of the lowest
eigen-potential surface, $E_{-}(\beta ,\gamma )$, for
$^{98}$Zr, $^{100}$Zr and $^{102}$Zr. Adapted from~\cite{GavLevIac19}.}
\end{figure*}

A geometric visualization of the IBM-CM is obtained by introducing
coherent (intrinsic) states~\cite{gin80,diep80}
and constructing an energy-surface
matrix whose entries are the matrix elements of the
Hamiltonian~(\ref{Hmat}) between the intrinsic states of
the two configurations.
Diagonalization of this two-by-two
matrix produces the so-called eigen-potentials,
$E_{\pm}(\beta,\gamma)$~\cite{frank}.
In Figure~\ref{Eminus}, we show the calculated 
lowest eigen-potential $E_{-}(\beta ,\gamma)$
for the isotopes described in Figure~\ref{fig:spectrum}.
In general, these classical potentials confirm the quantum results,
as they show a transition from spherical ($^{92-98}$Zr),
to a flat-bottomed potential at $^{100}$Zr,
to prolate axially-deformed ($^{102-104}$Zr), and finally to 
$\gamma$-unstable ($^{106-110}$Zr).

In general, the results of the current phenomenological
study resemble those
obtained in the microscopic approach
of the MCSM~\cite{taka}
(which focuses on spectra and $E2$ rates), 
however, there are some noticeable differences. Specifically,  
the replacement $\gamma$-unstable $\rightarrow$ triaxial 
and the inclusion of more than two configurations in the MCSM. 
The spherical state in $^{100}$Zr is identified in the MCSM as $0^{+}_4$,
in contrast to $0^{+}_2$ in the current calculation and the data.
Both calculations show a large jump in 
$B(E2;2^+_1\rightarrow0^+_1)$, between $ ^{98} $Zr and $ ^{100} $Zr, 
typical of a first-order QPT. This is in contrast with mean-field 
based calculations~\cite{delaroche,nomura16,mei},
which due to their character smooth out the phase transitional
behavior, and show no such jump at the critical point of the QPT
(see Figure~2 of~\cite{Singh18}).
The observed peak in $B(E2;2^+_1\rightarrow0^+_1)$ for $^{104}$Zr,
is reproduced by the current calculation but not by the MCSM.

\section*{Acknowledgements}
This work was supported in part by the US-Israel Binational Science
Foundation Grant No. 2016032 and by the U.S. DOE under Grant No.
DE-FG02-91ER-40608. We thank R.~F. Casten and J.~E.~Garc\'\i a-Ramos for
insightful discussions.\\

\end{document}